\documentclass[10pt]{article}

\usepackage{amssymb}

\usepackage[margin=1in]{geometry}  
\usepackage{graphicx}              
\usepackage{amsmath}               
\usepackage{amsfonts}              
\usepackage{amsthm}                
\usepackage{caption}
\usepackage{subcaption}
\usepackage{array}
\usepackage[usenames]{color}
\usepackage{colortbl}
\usepackage{xcolor}
\usepackage[utf8]{inputenc}
\usepackage[english]{babel}
\usepackage[document]{ragged2e}
\usepackage{epstopdf}
\usepackage{amsmath}
\usepackage{graphicx}
\usepackage{pict2e}

\usepackage{bm}
\newcommand\abs[1]{\left|#1\right|}

\begin{document}



\title{Linear Theory Analysis of Self-Amplified Parametric X-ray Radiation from High Current Density Electron Bunches}
\maketitle

\center{\author{Ihar Lobach, Andrei Benediktovitch and Ilya Feranchuk}}

Physics Department,  Belarusian State University,
 4, Nezavisimosti Av., 220030 Minsk, Belarus
\justify
\begin{abstract}
Linear theory of the parametric beam instability or  the self-amplification of parametric x-ray radiation (PXR) from relativistic electrons in  a crystal is considered taking into account   finite emittance of the electron beam and absorption of the radiation. It is shown that these factors change essentially the estimation of  threshold parameters of the electron bunches for the coherent X-ray generation. The boundary conditions for the linear theory of the effect is analyzed in details and it is shown that  the grazing incidence diffraction geometry is optimal for the growth of instability. Numerical estimations of amplification and coherent photon yield in dependence on the electron current density are presented for the case of mm-thickness Si crystal   and 100 MeV electrons. Possible improvements of the experimental scheme for  optimization of the coherent radiation intensity are discussed.
\end{abstract}

\section{Introduction}
The advent of X-ray free electron lasers (XFEL) opened new era in the investigation of matter on Angstrom lengthscale with fs time resolution \cite{2011_xfel_chapman2011femtosecond,2012_xfel_biology,2014_xfel_barends2014novo,2014_xfel_xu2014single}. However, need for GeV electron accelerators and hundreds meter long undulator modules results in high construction and maintenance costs. At the moment there are two operating facilities \cite{2010_lcls_emma2010first,2012_sacla_ishikawa2012compact} which are extremely overbooked and several projects under construction \cite{2007_euroxfel_altarelli2007euroxfel,2010_xfel_list_mcneil2010x}. In recent years there were investigated several mechanisms that could lead to compact lab-size bright and coherent x-ray sources \cite{18,19,20,21}. In this contribution we will theoretically analyze the possibility to achieve the x-ray lasing from 100 MeV electrons in mm-thick crystals based on the parametric beam instability effect. This effect was predicted by Baryshevsky and Feranchuk in 1983 \cite{17}, they showed that above threshold current density value the interaction between parametric x-ray radiation electric field and relativistic electrons leads to instability and exponential growth of radiated intensity. This effect was realized in the THz range with artificial periodic structures \cite{2002_vfel_baryshevsky2002first}, but for x-rays and crystals the threshold current density was estimated to be $10^{9}A/cm^{2}$. Such current density values were considered to be unrealistic at the time of publication \cite{17}, but became recently available from short electron bunches. It is also important to stress that such bunches could pass through the mm-thickness crystal without its destruction during the production of x-ray radiation \cite{PBI_SPIE}, the concept being close to "diffraction before destruction" one, used intensively in the XFEL imaging approaches \cite{2014_chapman2014diffraction}.

Before quantitative estimations, let us qualitatively describe what we expect of beam instability. In case of incoherent PXR the intensity of radiation is proportional to the number of electrons in the beam, as they are arranged randomly in the beam. However, emitted field affects trajectories of the electrons, this process may be called self-interaction. It may happen that rearrangement of the electrons takes place and they become ordered in space due to self-interaction. In that case electrons start to radiate coherently and one can expect radiation intensity to grow up to proportional to squared number of electrons in the beam. For a nC bunch this may result in $10^{10}$ increase of intensity compared to spontaneous PXR case, the origin of this enhancement is the same as SASE process in XFEL. The aim of the present contribution is to investigate the parameters of the electron beam and experimental geometry under which the coherent amplification starts.

In the present contribution we use the classical electrodynamics and linearized theory that enable to describe the initail phase of intensity amplification and locate the necessary for the parametric beam instability parameter area. We take into account absorption in the crystal, consider growth of instability out of incoherent PXR under grazing incidence geometry and estimate influence of the electron beam's emittance. The paper is organized as follows. The theoretical basics is developed in Sec.\ref{sec:theory}. The development is done in following steps: In Sec.\ref{sec:hpoebc} we calculate the the electric current originating from an external filed. It can be represented as $\bm{\widehat{J}E}$ (in $\bm{k}-w$ domain), where $\bm{\widehat{J}}$ is a $3\times 3$ matrix,it is the operator $\bm{\widehat{J}}$ that describes self-interaction. In Sec.\ref{sec:dispeq} we derive the dispersion equation in case of beam instability in order to be able to find the field eigenwaves in the crystal. Sec.\ref{sec:intersection} describes how to pick out the most amplified waves. In Sec.\ref{sec:BC} we analyze the boundary conditions and interplay between spontaneous and amplified fields. Based on the analysis, the amplitudes of out-coming waves are calculated. Sec.\ref{sec:sne} provides several numerical examples. In Sec.\ref{sec:discussion} the obtained results are analyzed and ways for improvement are outlined.

\section{Calculations}\label{sec:theory}
\subsection{Notations}
\begin{figure}
  \centering
  \includegraphics[width=0.4\textwidth]{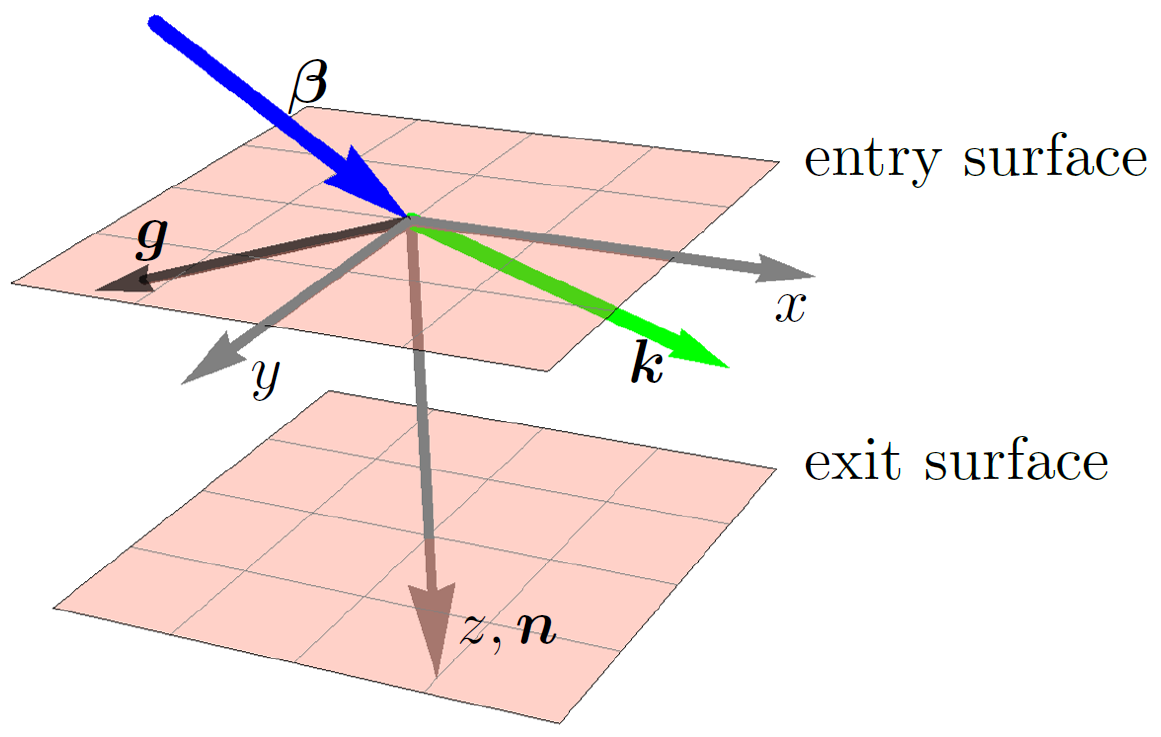}\\
  \caption{Coordinate system conventions.}\label{fig:figure7}
\end{figure}

In present paper we make use of Gaussian-cgs units. Further, we try to avoid frequency $w$ itself and use the combination $w/c$ instead. Hereafter we will imply $w/c$ by $w$ in order to shorten equations. The beam electrons' velocity is described by dimensionless value $\bm{\beta}=\bm{v}/c$. The coordinate system we use is shown if Fig.\ref{fig:figure7}.

\noindent $\bm{n}$ is a surface normal, $\abs{\bm{n}}=1$, $\bm{g}$ is a vector of reciprocal lattice space, it is parallel to the crystal's surface as we consider grazing incidence geometry. In this problem we will have to deal with two strong waves $\bm{k}$ and $\bm{k_{g}}=\bm{k}+\bm{g}$. $\bm{k}$ will either lie in $xz$-plane or be close to it.
\subsection{Homogeneous part of the electron beam current}\label{sec:hpoebc}
In this subsection we consider two forms of the beam current density, they correspond to zero and nonzero emittance of the electron beam, in case of nonzero emittance we regard distribution over directions of electrons' velocities in the beam only. Although distribution over magnitudes of velocities has influence on the effect, impact of distribution over directions is more significant. Let us first consider the zero emittance case. The beam is assumed to be uniform. We represent coordinate and velocity of an electron of the beam numbered $j$ as
\begin{align}
\begin{aligned}
&\bm{\beta}_{j}(t)=\bm{\beta}+\bm{\delta \beta}_{j}(t)\\
&\bm{r}_{j}(t)=\bm{r}_{j0}+\bm{\beta} c t+\bm{\delta r}(t)\\
&\bm{\dot{\delta r}}_{j}(t)=c\bm{\delta\beta}_{j}(t)
\end{aligned}
\end{align}
$\bm{\delta \beta}_{j}(t)$ is obtained from the differential equation which is consequence of the second Newton's law \cite{1}:
\begin{align}\label{eq:SNL}
 \bm{\dot{\delta\beta}}_{j}=\frac{-e}{m c \gamma_{i}}((\bm{1}-\bm{\beta}_{j}\bm{\otimes \beta}_{j})\bm{E}(\bm{r}_{j},t)+\bm{\beta}_{j}\bm{\times H}(\bm{r}_{j},t)),
\end{align}
\noindent here $\gamma_{j}=1/\sqrt{1-\beta_{j}^{2}}$, $\bm{\beta_{j}}$,$\bm{\delta \beta}_{j}$ and $\bm{r}_{j}$ are functions of time, $m$ is an electron's mass, $\otimes$ denotes the outer (Kronecker) product. In the framework of linear approximation $\bm{\delta \beta_{j}}$ is found from Eq.(\ref{eq:SNL}) under substitution $\bm{\beta}_{j}\rightarrow\bm{\beta}$ in the right-hand side. Then, the homogeneous part of the beam current can be found by the following expression

\begin{align}
\bm{j}(\bm{r},t)=-e c\sum_{j}\bm{\delta \beta}_{j}(t)\delta\left(\bm{r}-\bm{r}_{j}(t)\right)
\end{align}

 In this approximation after routine calculations one obtains

\begin{align}\label{eq:eqj}
\bm{j}(\bm{k},w)=\frac{i e^{2} n_{b}}{m c \gamma w}(\bm{1}+\frac{\bm{\beta \otimes k}+\bm{k \otimes \beta}}{w-\bm{k\cdot\beta}}+\frac{k^{2}-w^{2}}{(w-\bm{k\cdot \beta})^{2}}\bm{\beta\otimes\beta})\bm{E}(\bm{k},w)
\end{align}

\noindent here $n_{B}$ - is the concentration of electrons in the beam ($1/m^{3}$),

\begin{align}
\bm{j}(\bm{k},w)=\int \bm{j}(\bm{r},t)\exp{i(w t-\bm{k}\cdot\bm{r})}\bm{dr}dt
\end{align}

The approach to calculation of the formula (\ref{eq:eqj}) is described in \cite[pp.146-153]{1} and \cite{4} in detail. Also, Eq.(\ref{eq:eqj}) can be derived by a relativistic approach like in \cite[pp.144-145]{5}. One can obtain an expression for current in a frame of reference, where mean velocity of electrons of the beam equals zero, and then transfrom into the laboratory frame of reference by using Lorentz transformations.

In present research we consider the vicinity of Cherenkov's resonance, that means $\abs{w-\bm{k\cdot \beta}}<<w$. It lets us omit first and second addends in Eq.(\ref{eq:eqj}). Furthermore, for the wave $\bm{k_{g}}$ we can neglect the whole beam current by the same reasoning.

\begin{figure}
  \centering
  \includegraphics[width=0.4\textwidth]{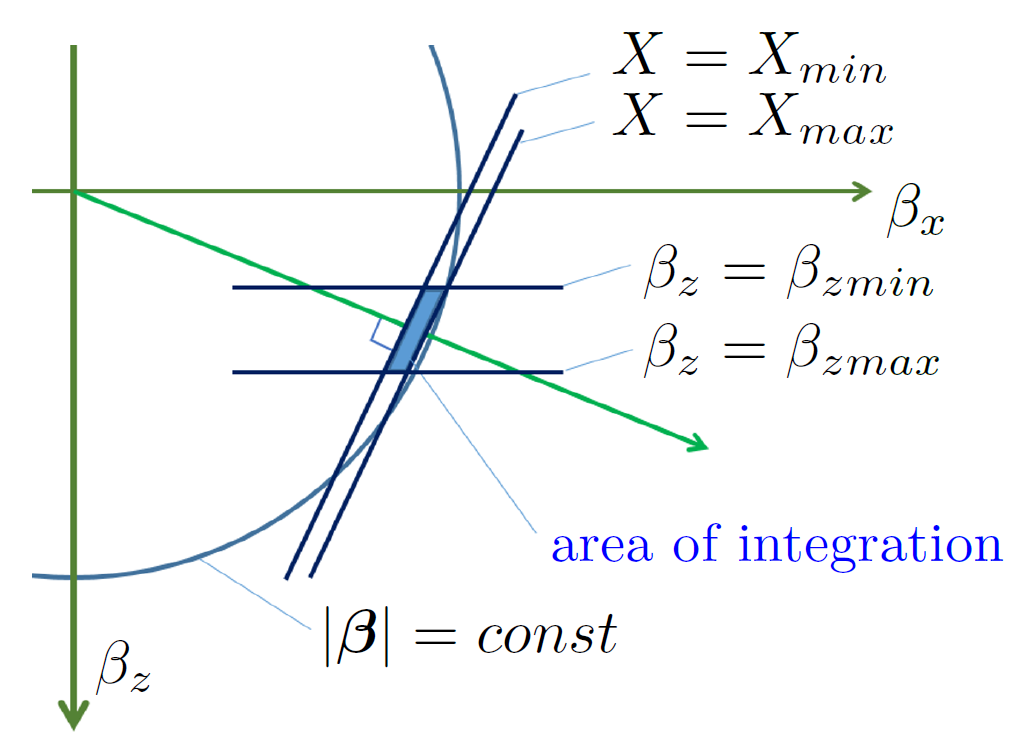}\\
  \caption{Integration area in coordinates $\beta_{x}$, $\beta_{z}$. $\bm{k_{0}}$ lies in $\beta_{x}\beta_{z}$ plane. In more details, the area is placed on the surface of a sphere $\abs{\bm{\beta}}=const$ and the figure represents a projection on the plane $\beta_{x}\beta_{z}$.}\label{fig:figure1}
\end{figure}

As to nonzero emittance case, then we make use of the approximate expression for the current and take into account distribution over directions of velocities by performing an integration

\begin{align}\label{eq:eq10}
\bm{j}(\bm{k},w)=\frac{i e^{2}}{m c \gamma  w}\int\frac{dn_{B}}{\bm{d\beta}}\frac{k^{2}-w^{2}}{(w-\bm{k\cdot \beta})^{2}}\bm{\beta\otimes\beta}\bm{d\beta}\bm{E}(\bm{k},w),
\end{align}

\noindent here $\abs{\bm{\beta}}$ in the integration area is implied to be constant. Before we choose the integration region we have to introduce a vector $\bm{k_{0}}$, which is an initial approximation to the solution of dispersion equation so that $\bm{k}$ can be represented as $\bm{k}=\bm{k_{0}}+\bm{n}\delta_{z}$, where $\abs{\delta_{z}}<<w$. In this case $w-\bm{k\cdot \beta}=(w-\bm{k_{0}\cdot \beta})-\beta_{z}\delta_{z}$. The form of the velocity distribution does not change the qualitative behavior of the function (\ref{eq:eq10}), and we will choose the from which makes the equations easier. It was convenient for us to choose a rectangular integration area in parameters $(w-\bm{k_{0}\cdot \beta})$, $\beta_{z}$. We denote $X=-(w-\bm{k_{0}\cdot \beta})$ for compactness. Also, we assume that $dn_{B}/(dXd\beta_{z})=const$ in the integration region. Visualization of such integration area is given in Fig.\ref{fig:figure1} . We chose this particular distribution because it can be integrated analytically. However, it is realizable as it does not have any singularities or other peculiarities except for its abrupt border. Still, one can consider it as a model of real distribution with realistically sharp border.

Although exact expression for the integral in Eq.(\ref{eq:eq10}) was found, it is quite cumbersome and there is no need to present it here. We are much more interested in asymptotic behavior of this integral at $\delta_{z}\rightarrow 0$. Concerning this, it was established that at arbitrary limits of integration over $X$ the integral tends to a constant when $\delta_{z}\rightarrow 0$, but if either $X_{min}$ or $X_{max}$ equals zero, the integral acts like $(1/\delta_{z})\bm{\widehat{B}}$ when $\delta_{z}\rightarrow 0$, where $\bm{\widehat{B}}$ is some operator independent on $\delta_{z}$. From now on we consider the case $X_{min}=0$, even though results are essentially the same in either case. Here is an expression for $\bm{\widehat{B}}$
\begin{align}
\bm{\widehat{B}}=\frac{i e^{2} n_{B}}{m c \gamma w}\frac{k^{2}-w^{2}}{X_{max}}\begin{pmatrix}
B_{11}&0&B_{13}\\
0&B_{22}&0\\
B_{31}&0&B_{33},\\
\end{pmatrix}
\end{align}

where

\begin{align*}
\begin{aligned}
&B_{11}=\frac{1}{k_{0x}^{2}}\left(k_{0z}(-2w+\frac{1}{2}k_{0z}(\beta_{zmin}+\beta_{zmax}))+w^{2}\frac{\ln\beta_{zmax}-\ln\beta_{zmin}}{\beta_{zmax}-\beta_{zmin}}\right)\\
&B_{22}=\frac{1}{k_{0x}^{2}}\left(2k_{0z}w-\frac{1}{2}k_{0}^{2}(\beta_{zmin}+\beta_{zmax})-(w^{2}-k_{0x}^{2}\beta^{2})\frac{\ln\beta_{zmax}-\ln\beta_{zmin}}{\beta_{zmax}-\beta_{zmin}}\right)\\
&B_{33}=\frac{1}{2}(\beta_{zmax}+\beta_{zmin})\qquad B_{31}=B_{13}=\frac{1}{k_{0x}}(w-\frac{1}{2}k_{0z}(\beta_{zmin}+\beta_{zmax}))\\
&n_{b}=\int\frac{dn_{b}}{\bm{d\beta}}\bm{d\beta}
\end{aligned}
\end{align*}

$k_{0y}=0$, for coordinate system see Fig.\ref{fig:figure7}. Fortunately, it turns out that under certain conditions the operator $\bm{\widehat{B}}$ can be represented in a form of a diad proportional to $\bm{\widetilde{\beta}\otimes \widetilde{\beta}}$ with a relative error $\sim 10^{-3}$, where $\bm{\widetilde{\beta}}$ is one from such $\bm{\beta}$ that $w-\bm{k_{0}\cdot\beta}=0$. We'll consider such configurations only. Finally, coming back to universal notation, one can state that in nonzero emittance case

\begin{align}\label{eq:j_beta_otimes_beta}
\bm{j}(\bm{k},w)\sim\frac{\bm{\beta\otimes\beta}}{(w-\bm{k\cdot \beta})}\bm{E}(\bm{k},w)
\end{align}

\subsection{Dispersion equation}\label{sec:dispeq}
In the framework of two-wave approximation Maxwell's equations lead to the following linear system(see \cite[pp.146-153]{1}), where we have already omitted negligible parts of the current:

\begin{align}\label{eq:eqs1}
\begin{aligned}
&(T-\bm{k\otimes k})\bm{E}-w^{2}\chi_{-g}\bm{E_{g}}=-Y\bm{\beta\otimes\beta}\bm{E}\\
&(T_{g}-\bm{k_{g}\otimes k_{g}})\bm{E_{g}}-w^2\chi_{g}\bm{E}=0,
\end{aligned}
\end{align}

where

\begin{align}\label{eq:eqY}
Y=\frac{w_{b}^{2}}{\gamma}(k^{2}-w^{2})\cdot
\begin{cases}
\frac{1}{(w-\bm{k\cdot\beta})^{2}}&\textbf{zero emittance case}\\
\frac{C}{(w-\bm{k\cdot\beta})}&\textbf{nonzero emittance case}
\end{cases}
\end{align}

\noindent here $w_{b}^{2}=4\pi e^{2}n_{b}/(m c^{2})$, $C$ is a coefficient dependent on parameters of distribution of  electrons' velocities over directions. $\bm{k_{g}}=\bm{k}+\bm{g}$, $\bm{E}=\bm{E}(\bm{k},w)$, $\bm{E_{g}}=\bm{E}(\bm{k_{g}},w)$, $\epsilon_{0}$ is a mean susceptibility of the crystal, $\chi_{g},\chi_{-g}$ are Fourier components of the crystal's susceptibility, $T=k^{2}-w^{2}\epsilon_{0}$, $T_{g}=k_{g}^2-w^{2}\epsilon_{0}$.

In general, there is a straightforward way to obtain the dispersion equation, it is to find the determinant of the $6\times 6$ matrix corresponding to the system of equations (\ref{eq:eqs1}) and equate it to zero. One may even take more accurate expression for current. However, here we present more demonstrative derivation of approximate dispersion equation for $\sigma$-polarization in case of zero emittance which proves to give solutions almost identical to those of the straightforward dispersion equation.

Let us regard such $\bm{\beta}$ that can be presented as
$$\bm{\beta}=\frac{\beta_{\parallel}}{k}\bm{k}+\bm{\beta_{\perp}}$$
with appropriate accuracy. Where $\bm{\beta_{\parallel}}=\left(\bm{\beta\cdot k}\right)/k$, $\bm{\beta_{\perp}\parallel\bm{k\times k_{g}}}\parallel\bm{e_{\sigma}}$ - the $\sigma$-polarization unit vector.

In this case
$$\bm{\beta\cdot E}=\frac{\beta_{\parallel}}{k}\bm{E\cdot k}+\beta_{\perp}E^{\sigma}$$
Scalar products of Eqs.(\ref{eq:eqs1}) and $\bm{e_{\sigma}}$ give

\begin{align}\label{eq:eqs2}
\begin{aligned}
&T E^{\sigma}-w^{2}\chi_{-g}E^{\sigma}_{g}=-\beta_{\perp}Y\left(\frac{\beta_{\parallel}}{k}\bm{E\cdot k}+\beta_{\perp}E^{\sigma}\right)\\
&T_{g}E_{g}^{\sigma}-w^{2}\chi_{g}E^{\sigma}=0
\end{aligned}
\end{align}

We don't know $\bm{E\cdot k}$ yet. Scalar products of Eqs.(\ref{eq:eqs1}) and $\bm{k,k_{g}}$ give

\begin{align}
\begin{aligned}
\begin{cases}
&\left( T-k^{2}\right) \left(\bm{E\cdot k}\right)-w^{2}\chi_{-g}\left( \bm{E_{g}\cdot k}\right)=-\left( \bm{\beta \cdot k}\right) Y\left(\frac{\beta_{\parallel}}{k}\left(\bm{E\cdot k}\right)+\beta_{\perp}E^{\sigma}\right)\\
&T\left(\bm{E\cdot k_{g}}\right)-\left(\bm{k\cdot k_{g}}\right)\left(\bm{E\cdot k}\right)-w^{2} \chi_{-g}\left(\bm{E_{g}\cdot k_{g}}\right)=-\left(\bm{\beta\cdot k_{g}}\right) Y\left(\frac{\beta_{\parallel}}{k}\left(\bm{E\cdot k}\right)+\beta_{\perp}E^{\sigma}\right)\\
&T_{g}\left(\bm{E_{g}\cdot k}\right)-\left(\bm{k\cdot k_{g}}\right)\left(\bm{E_{g}\cdot k_{g}}\right)-w^2 \chi_{g}\left(\bm{E\cdot k}\right)=0\\
&\left(T_{g}-k_{g}^2\right)\left(\bm{E_{g}\cdot k_{g}}\right)-w^2 \chi_{g}\left(\bm{E\cdot k}\right)=0
\end{cases}
\end{aligned}
\end{align}

This is a linear system of equations for the variables $\left(\bm{E\cdot k}\right),\left(\bm{E\cdot k_{g}}\right),\left(\bm{E_{g}\cdot k}\right),\left(\bm{E_{g}\cdot k_{g}}\right)$. Thus, one finds $\left(\bm{E\cdot k}\right)$ through $E^{\sigma}$ and can substitute it into Eq.(\ref{eq:eqs2}). The latter becomes a homogeneous linear system for $E^{\sigma}$ and $E_{g}^{\sigma}$. In order that it has non-trivial solutions the determinant of the corresponding matrix should be equal to zero, which yields the following dispersion equation for $\sigma$-polarized waves

\begin{align} \label{eq:dispeq}
\left(w-\bm{k\cdot \beta}\right)^{2}D_{\sigma}=\frac{w_{b}^{2}}{\gamma w^{2}}\left(k^{2}-w^{2}\right)\left(D_{\sigma}-\beta_{\perp}^{2}w^{2}T_{g}\right),
\end{align}

where $ D_{\sigma}=T T_{g}-\chi_{g}\chi_{-g}w^{4}$ - left-hand side of XRD(X-ray diffraction) $\sigma$-polarization dispersion equation, here the zero-emittance case was considered. Eq.(\ref{eq:dispeq}) is valid in the vicinity of intersection of Cherenkov roots and one from diffraction roots. The addend $D_{\sigma}$ on the right-hand side of Eq.(\ref{eq:dispeq}) is due to non-zero $\bm{E\cdot k}$. It was verified that its role is significant. Hence, the longitudinal component of the field cannot be neglected and the approximation of transverse waves fails.

Eq.(\ref{eq:dispeq}) can help to get insight into the situation, to find a range of better parameters etc. In order to obtain an exact dispersion equation corresponding to the linear system (\ref{eq:eqs1}) we used the following technique. Substituting $\bm{E_{g}}$ from the first equation of Eqs.(\ref{eq:eqs1}) into the second one we get
\begin{align}
\left[(T_{g}-\bm{k_{g}\otimes k_{g}})((T-\bm{k\otimes k})+Y\bm{\beta\otimes\beta})-w^{4}\chi_{g}\chi_{-g}\right]\bm{E}=0
\end{align}
Scalar products of this equation with vectors $\bm{k},\bm{k_{g}},\bm{\beta}$ results in a homogeneous linear system for the variables $(\bm{E\cdot k}),(\bm{E\cdot k_{g}}),(\bm{E\cdot \beta})$. If one equates the determinant of this linear system to zero, then one obtains the dispersion equation corresponding to Eqs.(\ref{eq:eqs1}). From our viewpoint this approach is preferable, because it does not make us deal with any coordinate systems. Resultant expression contains scalar products of the vectors $\bm{k},\bm{k_{g}},\bm{\beta}$ only. Obtained dispersion equation is presented below. $Y$ is defined in Eq.(\ref{eq:eqY}).

\begin{align}\label{eq:eq11}
w^{4}D_{\sigma}D_{\pi}=Y\left[ D_{\sigma}\lbrace T_{g}w^{2}\epsilon_{0}(G^{2}-w^{2}\epsilon_{0}\beta^{2})-W(M^{2}-T_{g}\beta^{2})\rbrace-V^{2}T_{g}W\right],
\end{align}

where $G=\bm{k\cdot\beta}$, $M=\bm{k_{g}\cdot \beta}$, $V=\bm{k\cdot}(\bm{k_{g}\times \beta})$, $W=\chi_{g}\chi_{-g}w^{4}$,

\begin{align*}
D_{\pi}=(\epsilon_{0}^{2}-\chi_{g}\chi_{-g})\left[T T_{g}-W\left(1-\frac{k^{2}k_{g}^{2}-(\bm{k\cdot k_{g}})^{2}}{w^{4}(\epsilon_{0}^{2}-\chi_{g}\chi_{-g})}\right)\right]
\end{align*}

- left-hand side of the XRD dispersion equation for $\pi$-polarized waves.

If in Eq.(\ref{eq:eq11}) right-hand side equaled 0, it would be the XRD dispersion equation for total electric field. It would yield 8 diffraction roots. In considered case of beam instability $(w-\bm{k\cdot \beta})^{2}$ or $(w-\bm{k\cdot \beta})$ in denominator of $Y$ increases order of the equation by 2 or 1 respectively. The diffraction roots remain almost unchanged, additional 2 or 1 roots are by few orders of magnitude less than the diffraction roots in our parametrization, we will call them Cherenkov roots. Here by roots we mean $\delta_{z}$ in $\bm{k}=\bm{k_{0}}+\bm{n}\delta_{z}$. We will number all roots of Eq.(\ref{eq:eq11}) $\delta_{z}^{(m)}$ by $m=1..10$ or $1..9$. $m=1..8$ correspond to diffraction roots, the rest - Cherenkov roots. Until this moment we distinguished the two forms of current by corresponding beam emittance. Now we see, that we might as well call them two- and one Cherenkov roots cases. It will be used further in Sec.\ref{sec:BC}.

\subsection{Intersection of diffraction and Cherenkov roots}\label{sec:intersection}
\begin{figure}
  \centering
  \includegraphics[width=0.4\textwidth]{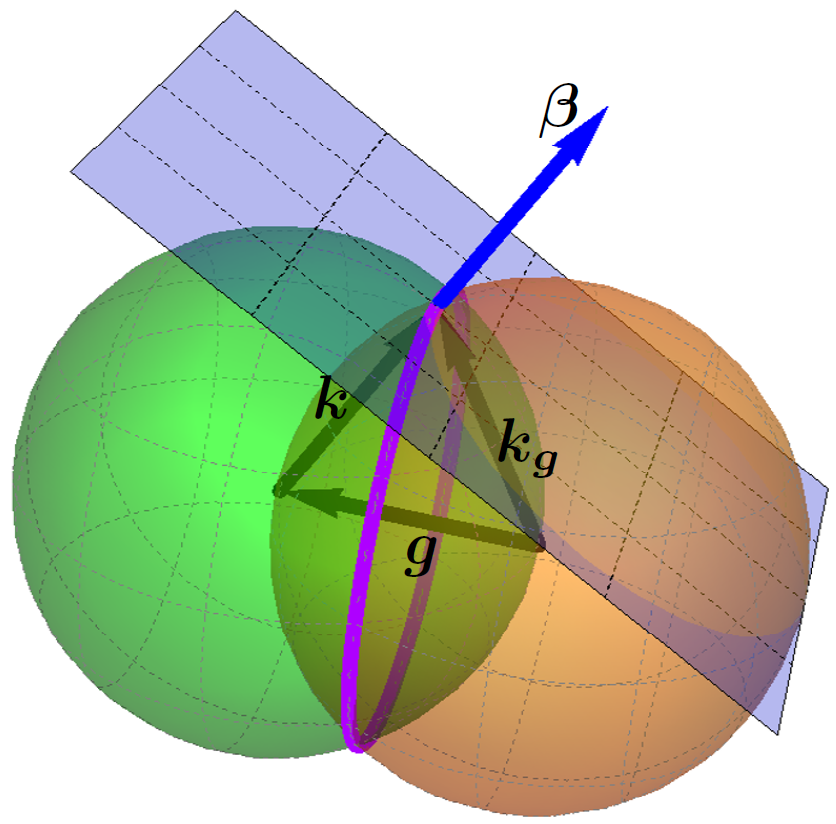}\\
  \caption{Intersection of dispersion surfaces. Green sphere - $\abs{\bm{k}}=w$, brown sphere - $\abs{\bm{k_{g}}}=w$, blue plane - $w-\bm{k\cdot \beta}=0$ - the Cherenkov plane. It is supposed that $\bm{\beta}=\bm{const}$ and $w=const$ here. Thick  purple line denotes the points where Bragg's condition is fulfilled.}\label{fig:figure2}
\end{figure}

Our aim is to maximize the increment of parametric beam instability. Hence, we have to choose an appropriate region in $\bm{k}-w$ for investigation. It corresponds to intersection of Cherenkov and one from diffraction roots. First approximation to solution of this problem is shown in Fig.\ref{fig:figure2}.

%

The surfaces in Fig.\ref{fig:figure2} are situated in a vector space of $\bm{k},\bm{k_{g}},\bm{g}$. The two spheres are the same as in the problem of two wave dynamical diffraction theory (see \cite[p.131]{7}). The purple line corresponds to fulfillment of Bragg's condition(see \cite[p.121]{7}) and noticeable interactions of the waves $\bm{k}$ and $\bm{k_{g}}$. Apart from this intersection in case of bunch instability there should be an intersection of this line with the Cherenkov plane. To be precise, there is no exact intersection with the Cherenkov plane in Fig.\ref{fig:figure2}, we can only speak of the point of the plane closest to the purple line. However, as already said, it is only first approximation, we will deal with the vicinity of this "intersection" in more detail below.

The region of parameters under investigation can be formulated in other words: Bragg's condition ($2\bm{k\cdot g}+g^{2}=0$) should be approximately fulfilled and the angle between $\bm{k}$ and $\bm{\beta}$ should be small.

Need for intersection of roots is seen in Eq.(\ref{eq:dispeq}). The term $(w-\bm{k\cdot \beta})$ and , consequently, the increment is larger as the term $D_{\sigma}$ is smaller.

Let us discuss how to find more exact intersection and first approximation to solution of the dispersion equation $\bm{k_{0}}$ (the solution is $\bm{k}=\bm{k_{0}}+\bm{n}\delta_{z}$). We have to consider $\bm{k_{0}}$ with real components only in order to satisfy intersection with the  Cherenkov plane $w-\bm{k_{0}\cdot \beta}=0$. We introduce an angle $\theta$ between $\bm{k_{0}}$ and $\bm{\beta}$  and rewrite last equation as

\begin{align}\label{eq:eq12}
w=k_{0}\beta\cos{\theta},
\end{align}

\noindent from now on let us consider $k_{0}$ and $\theta$ as free parameters, then Eq.(\ref{eq:eq12}) sets the value of $w$.

Further, in order to achieve intersection with one from diffraction roots, $\bm{k_{0}}$ should be a root of either $D_{\sigma}$ or $D_{\pi}$. We chose the former case. That is

\begin{align}\label{eq:eq14}
(k_{0}^{2}-w^{2}\epsilon_{0})(k_{0g}^{2}-w^{2}\epsilon_{0})-\chi_{g}\chi_{-g}w^{4}=0,
\end{align}

\noindent where $w$ is calculated by Eq.(\ref{eq:eq12}). One can represent $k_{0g}^{2}$ in the following way

\begin{align}\label{eq:eq13}
k_{0g}^{2}=k_{0}^{2}+2 \bm{k_{0}\cdot g}+g^{2}=k_{0}^{2}+k_{0}^{2}d,
\end{align}

\noindent where $d$ describes deviation from Bragg's  condition.

Owing to Eqs.(\ref{eq:eq12}) and (\ref{eq:eq13}) Eq.(\ref{eq:eq14}) turns into

\begin{align}\label{eq:eq15}
(1-\epsilon_{0}\cos^{2}\theta )(1+d-\epsilon_{0}\cos^{2}\theta )-\chi_{g}\chi_{-g}\cos^{4}\theta=0,
\end{align}

 \noindent which is a linear equation for $d$. However, exact $d$ obeying Eq.(\ref{eq:eq15}) is complex as $\epsilon_{0}$, $\chi_{g}\chi_{-g}$ are complex. It contradicts the statement that the components of $\bm{k_{0}}$ are real. There may be several ways to overcome this obstacle, we replaced $\epsilon_{0}$, $\chi_{g}\chi_{-g}$ in Eq.(\ref{eq:eq15}) with their real parts and solved it for $d$. Then, with good accuracy we can say that we are as close to intersection of diffraction and Cherenkov roots as possible.

\begin{figure}
  \centering
  \includegraphics[width=0.5\textwidth]{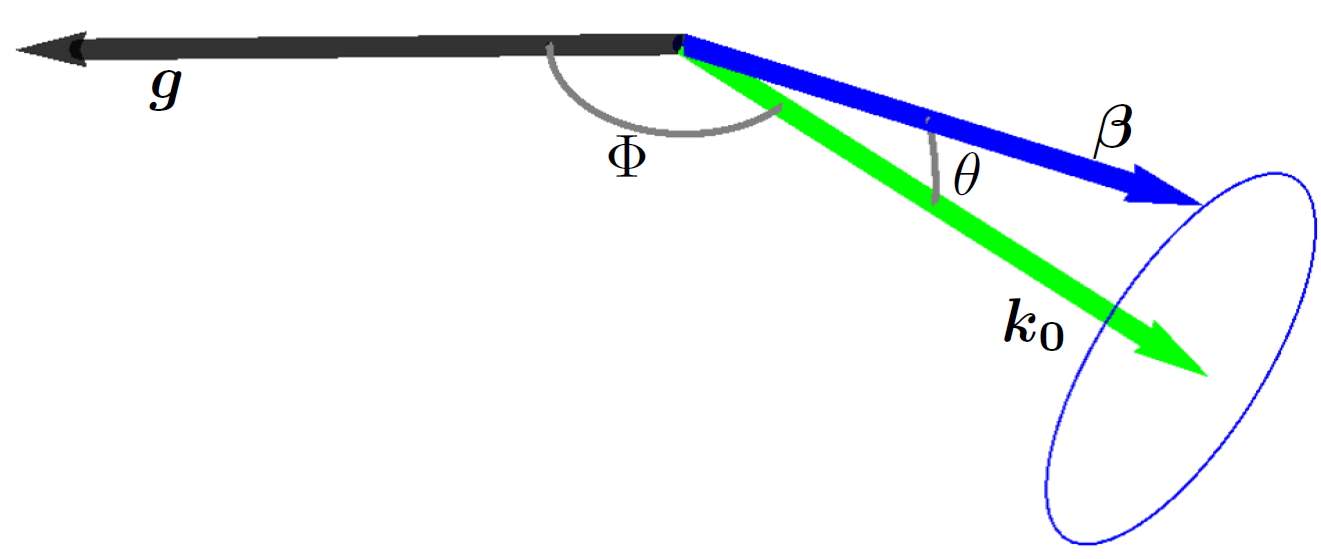}\\
  \caption{Relative orientation of $\bm{k_{0}}$, $\bm{g}$, $\bm{\beta}$ in case of intersection of one from diffraction roots and Cherenkov roots. The angle $\Phi$ is defined by $d$. The blue circle represents the freedom in choice of $\bm{\beta}$.}\label{fig:figure4}
\end{figure}

Having set values of $k_{0}$ and $\theta$, calculated by the above described procedure parameter $d$ sets relative orientation of $\bm{k_{0}}$ and $\bm{g}$, whereas the angle $\theta$ corresponds to relative orientation of $\bm{\beta}$ and $\bm{k_{0}}$, although there remains certain freedom in choice of orientation of $\bm{\beta}$ (it is shown in Fig.\ref{fig:figure4}). There is similar freedom in choice of $\bm{k_{0}}$ with respect to $\bm{g}$ as well.

%

Even though in the above paragraphs we tried to find solution of the problem step by step, now we can say that we merely have a system of equations relating $\bm{g},\bm{\beta}, \bm{k_{0}}$ and $w$ corresponding to the intersection. Hence, if one sets certain relative orientation of $\bm{g}, \bm{\beta}$ and $\bm{n}$ in experiment, then one can approximately know the most amplified out-coming waves.

It is necessary to note that if imaginary parts of $\epsilon_{0}$, $\chi_{g}\chi_{-g}$ vanished, then there would be exact intersection of one from diffraction roots and Cherenkov roots. In the vicinity of intersection $D_{\sigma}$ would be proportional to $(w-\bm{k\cdot \beta})$ or, in other words, to $\delta_{z}$.   Left-hand side of Eq.(\ref{eq:dispeq}) or (\ref{eq:eq11}) would be proportional to $\delta_{z}^{3}$ (in zero emittance case). When calculating Cherenkov $\delta_{z}$ from Eq.(\ref{eq:dispeq}) or (\ref{eq:eq11}) it is enough to retain the term $\delta_{z}^{3}$ and substitute $\delta_{z}\rightarrow 0$ in all other positions as Cherenkov $\delta_{z}$ is several orders of magnitude less than diffraction roots. Thereby, in case of zero imaginary parts of the crystal susceptibilities (see \cite[pp.146-153]{1}) Cherenkov $\delta_{z}\sim n_{b}^{1/3}\sim j^{1/3}$, where $j$ is the electron beam's current density. Clearly, the same behavior holds for increment. In present research we retain imaginary parts of $\epsilon_{0}$, $\chi_{g}\chi_{-g}$ and thereby take into account absorption. In this case in the vicinity of "intersection" $D_{\sigma}$ is not proportional to $\delta_{z}$, it is approximately constant, although the constant being quite small. That fact reduces order of $\delta_{z}$ in $\delta_{z}^{3}$ down to $\delta_{z}^{2}$ in zero emittance case and down to $\delta_{z}$ in nonzero emittance case. To sum up, under absorption in zero emittance case $\delta_{z}\sim j^{1/2}$, in nonzero emittance case $\delta_{z}\sim j$.

\subsection{Boundary problem}\label{sec:BC}
We use the following approach to estimation of the growth of instability. All but one electrons of the beam are considered as a medium with certain susceptibility, whereas the remaining electron is assumed to travel along the crystal unperturbed by the field, providing inhomogeneous part of the current. For more information on this method see \cite{3,2,10} and \cite[pp.377-402]{11}. The equations for the field become

\begin{align}\label{eq:eqs3}
\begin{aligned}
&(T-\bm{k\otimes k})\bm{E}-w^{2}\chi_{-g}\bm{E_{g}}=-Y\bm{\beta\otimes\beta}\bm{E}+H \bm{\beta}\delta\left(w-\bm{k\cdot\beta}\right) \\
&(T_{g}-\bm{k_{g}\otimes k_{g}})\bm{E_{g}}-w^2\chi_{g}\bm{E}=0,
\end{aligned}
\end{align}

\noindent here $H=8\pi^{2}i w e/c$, the addend with the Dirac delta function is the unperturbed electron's current, see for example \cite{2}. One can obtain equations for the field in case of incoherent PXR by substitution $Y\rightarrow 0$ in Eqs.(\ref{eq:eqs3}).

General solution of Eqs.(\ref{eq:eqs3}) is the sum of general solution of the homogeneous system of Eqs.(\ref{eq:eqs1}) and a particular solution of the inhomogeneous system of Eqs.(\ref{eq:eqs3}). In can be expressed like (see \cite{3})
\begin{align}
\bm{E}(\bm{k},w)=\bm{E^{i}}(\bm{k},w)+\sum_{m}\bm{E^{h}}_{m}(w)\delta\left(k_{n}-k_{n}^{(m)}(w)\right),
\end{align}

\noindent here $i$ and $h$ stand for in- and homogeneous respectively. $\bm{E^{i}}(\bm{k},w)$ is derived from Eqs.(\ref{eq:eqs3}) in assumption that $\delta\left(k_{n}-k_{n}^{(m)}(w)\right)$ is an ordinary variable, $k_{n}$ - normal to the surface component of the wave vector, $k_{n}^{(m)}$ - $m$-th solution of the dispersion equation for $k_{n}$. The directions of  $\bm{E^{h}}_{m}(w)$ are determined by Eqs.(\ref{eq:eqs1}) at corresponding $k_{n}^{(m)}$, while their magnitudes are arbitrary constants. One may apply similar consideration to the field in vacuum. Then the standard boundary conditions are (see \cite{2})

\begin{align}\label{eq:eq4}
\int \bm{E_{c}}(\bm{k_{\tau}+k_{n}^{c}},w)d k_{n}^{c}=\int \bm{E_{v}}(\bm{k_{\tau}+k_{n}^{v}},w)d k_{n}^{v},
\end{align}

\noindent here $c$ and $v$ stand for crystal and vacuum respectively, by $\bm{E}$ here we imply total field, the sum of in- and homogeneous contributions. In general, there may be any other continuous across the boundary function in place of $\bm{E}(\bm{k},w)$ in Eq.(\ref{eq:eq4}). It was proven that $\bm{E}$, its normal derivative, the electron beam charge density and current comply with Eq.(\ref{eq:eq4}) with a good accuracy.

In present contribution we consider grazing incidence geometry (see \cite[pp.154-160]{7}), such choice will be justified below. Under this geometry all the diffraction roots should be taken into account. In the case of diffraction only, there are 8 $\bm{k}$-waves and 8 $\bm{k_{g}}$-waves. By $\bm{k}$- or $\bm{k_{g}}$-waves here we mean waves with in-surface components of wave vector equal to those of $\bm{k_{0}}$ or $\bm{k_{0}+g}$ respectively. Amplitude of each $\bm{k_{g}}$-wave is unambiguously related to amplitude of one from $\bm{k}$-waves. Hence, one can speak of only 8 eigenwaves of the field in the crystal, half of the waves being direct waves and the other half being specularly reflected ones.

\begin{figure}
  \centering
  \includegraphics[width=0.6\textwidth]{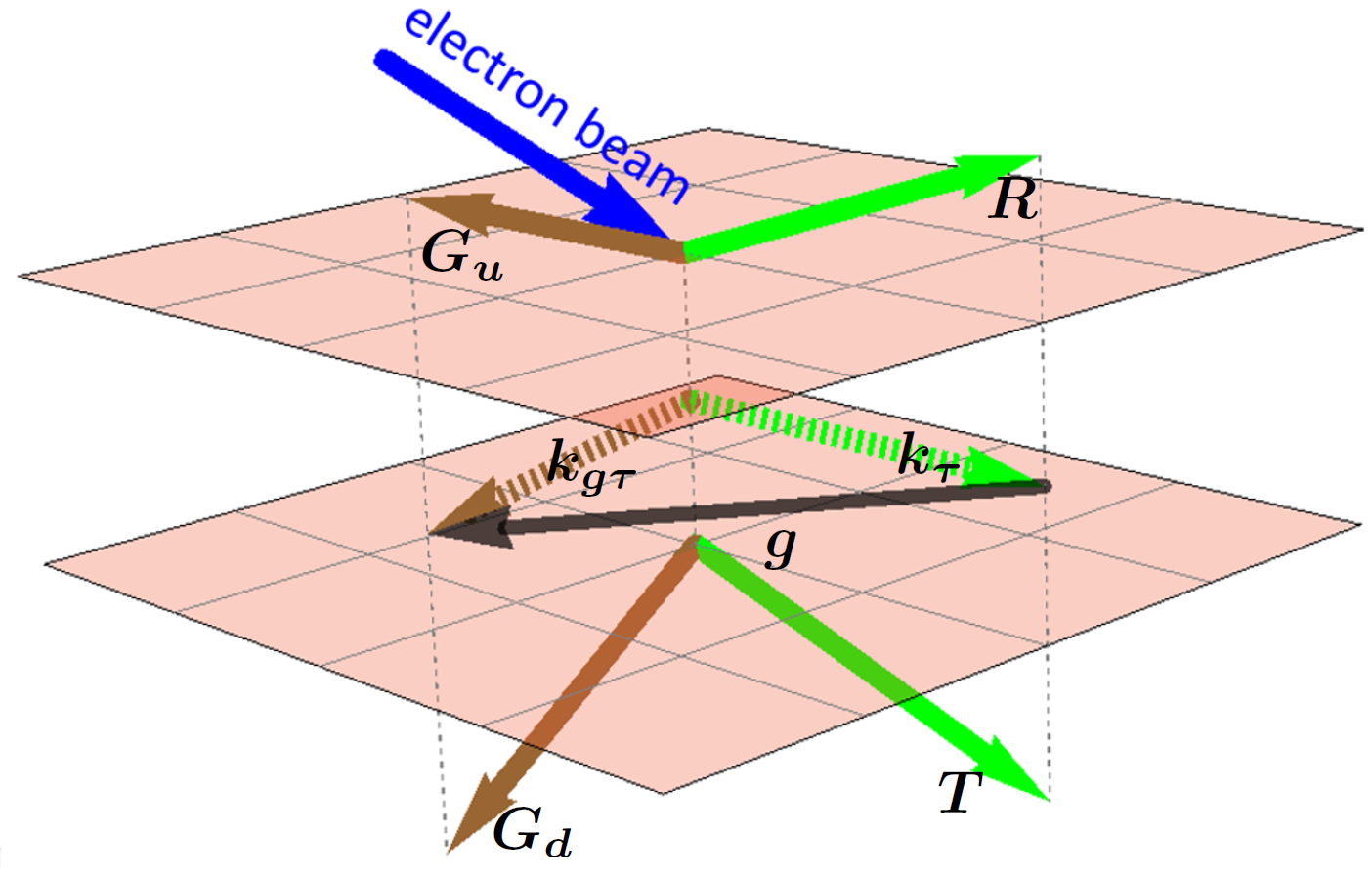}\\
  \caption{Sketch of the out-coming waves. $\bm{R},\bm{T},\bm{G_{u}},\bm{G_{d}}$ are amplitudes of the waves shown by green and brown arrows. $\bm{k_{\tau}}$ and $\bm{k_{g\tau}}$ are in-surface components of wave vectors of the out-coming waves.}\label{fig:figure3}
\end{figure}

As to the problem with beam instability, as in the previous case, the amplitudes of $\bm{k_{g}}$- and $\bm{k}$-waves are related for both diffraction and Cherenkov roots. Moreover, the relation remains the same, because the second equation of Eqs.(\ref{eq:eqs3}), relating the amplitudes, is identical to that in case of X-ray diffraction. Hence,  the unknowns are 10 or 9 (depending on the form of the beam current) scalar amplitudes of the eigenwaves and the vector amplitudes $\bm{R,T,G_{u},G_{d}}$ (actually, they have only two independent components as the field in vacuum is transverse)(see Fig.\ref{fig:figure3}). Summing up, one has 18 or 17 unknowns respectively. The straightforward way to formulate the boundary condition problem is to write down the equations of continuity of the field itself and its normal derivative on both boundaries of the crystal plate, moreover, one has to include into the system of equations the continuity of the beam current and charge density on the entry surface. More precisely, to get a well-conditioned linear system one should take the projections of the equations for the field and its derivative on corresponding polarization vectors from similar problem for incoherent PXR. It produces 16 equations. Further, in case of one Cherenkov root these equations should be supplemented by continuity of the beam current(although current density is a vector, the three equations are proportional to each other and only one independent equation is left). In case of two Cherenkov roots apart from the current continuity equation one can make use of continuity of the beam charge density. Hence, one has 18 and 17 independent equations for the cases of two and one Cherenkov roots respectively, which is enough to resolve corresponding boundary condition problems. Such consideration was performed and is exact in the framework of present research. Later it was noticed that there is much more demonstrative approximate solution to the formulated boundary problem. It undergoes only small deviations from the exact solution at  sufficiently low beam currents. Fortunately, such currents coincide with real electron beam currents. It is this approximate solution that is to be described below.

%

First, we'd like to state one from results of this approximate consideration. Let us deal with one Cherenkov root case and thick crystal plate($l>>L_{Bragg Ext.}$, here $L_{Bragg Ext.}$ is the secondary extinction length). Suppose one has a solution for the boundary condition problem for incoherent PXR. Now, if one wants to get $\bm{R,T,G_{u},G_{d}}$ in the case with bunch instability, then he only has to multiply $\bm{T}$ and $\bm{G_{d}}$ by $\exp(i\delta_{z}^{(9)}l)$ and leave $\bm{R}$ and $\bm{G_{u}}$ unchanged, $\delta_{z}^{(9)}$ is the Cherenkov root. Thus, it is obvious why there is an exponential increase in intensity of radiation from this perspective.

Let us prove it. Both in- and homogeneous contributions to the field are expressed through the Dirac delta functions and after integration in Eq.(\ref{eq:eq4}) there are nonsingular amplitudes left. From now on we will denote by $\bm{E}$ these integrated over $k_{n}$ amplitudes and will formulate the equations for them.

It is interesting that if one derives $\bm{E^{i}}$ from Eqs.(\ref{eq:eqs3}) then one finds that the inhomogeneous field itself vanishes. It was proven analytically for both variants of current. There were obtained expressions $\int f_{2}\cdot (w-\bm{k\cdot \beta})^{2}\delta(w-\bm{k\cdot \beta})dk_{n}$ and $\int f_{1}\cdot (w-\bm{k\cdot \beta})\delta(w-\bm{k\cdot \beta})dk_{n}$ for the cases of two Cherenkov roots and one Cherenkov root respectively, $f_{1}, f_{2}$ - certain functions with no singularities. However, in the equation for continuity of current  the contribution  due to the inhomogeneous field persists and equals exactly $H\bm{\beta}/\beta_{z}$, which is the integrated over $k_{n}$ current of the unperturbed by the field electron. The contributions are non-zero because they are calculated by expressions of the following form. $\int j_{2}\cdot (w-\bm{k\cdot \beta})^{2}/(w-\bm{k\cdot \beta})^{2}\delta(w-\bm{k\cdot \beta})dk_{n}$ and $\int j_{1}\cdot (w-\bm{k\cdot \beta})/(w-\bm{k\cdot \beta})\delta(w-\bm{k\cdot \beta})dk_{n}$ for the cases of two Cherenkov roots and one Cherenkov root respectively, $j_{1}, j_{2}$ - certain functions with no singularities. Thus, the contribution to the current due to the inhomogeneous field is in the origin of the whole effect.

Let us now consider the case of one Cherenkov root only. There is $\delta_{z}$ in the denominator of the expression for the beam current density. Magnitudes of the diffraction roots are much bigger than that of the Cherenkov root. It was verified by performing numeric calculations that the beam currents due to diffraction eigenwaves can be neglected. The above arguments let us finally write down the equation stating continuity of the electron beam current density
\begin{align}\label{eq:eq5}
\bm{\widehat{J}E_{9}}=\bm{J_{0}}
\end{align}
 ,where $\bm{\widehat{J}}=-Y\bm{\beta\otimes\beta}$, $\bm{\widehat{J}E_{9}}$ - the contribution due to the Cherenkov eigenwave , $\bm{J_{0}}=H\bm{\beta}/\beta_{z}$ - the contribution due to the inhomogeneous field. Eq.(\ref{eq:eq5}) doesn't contain amplitudes of diffraction eigenwaves. Hence, the Cherenkov eigenwave's amplitude can be found immediately through Eq.(\ref{eq:eq5}). The remaining diffraction amplitudes and $\bm{T,R,G_{u},G_{d}}$ can be found from equations of continuity of the field and its derivative
 \begin{align}\label{eq:eqs6}
 \begin{aligned}
 &\textbf{entry surface}  &&\textbf{exit surface}\\
 &\sum_{m}\bm{E}_{m}=\bm{R}  &&\sum_{m}\exp(i\delta_{z}^{(m)}l)\bm{E}_{m}=\bm{T}\\
 &\sum_{m}k_{z}^{(m)}\bm{E}_{m}=-w_{z}\bm{R}  &&\sum_{m}\exp(i\delta_{z}^{(m)}l)k_{z}^{(m)}\bm{E}_{m}=w_{z}\bm{T}\\
 &\sum_{m}\bm{\widehat{g}}_{m}\bm{E}_{m}=\bm{G_{u}} &&\sum_{m}\exp(i\delta_{z}^{(m)}l)\bm{\widehat{g}}_{m}\bm{E}_{m}=\bm{G_{d}}\\
 &\sum_{m}k_{z}^{(m)}\bm{\widehat{g}}_{m}\bm{E}_{m}=-w_{zg}\bm{G_{u}} &&\sum_{m}\exp(i\delta_{z}^{(m)}l)k_{z}^{(m)}\bm{\widehat{g}}_{m}\bm{E}_{m}=w_{zg}\bm{G_{d}}
 \end{aligned}
 \end{align}
 ,where $m=1..9$, $m=9$ corresponds to the Cherenkov root, the rest - diffraction roots, $w_{z}=\sqrt{w^{2}-k_{\tau}^{2}}$, $w_{zg}=\sqrt{w^{2}-k_{g\tau}^{2}}$, $\bm{\widehat{g}}_{m}=\chi_{g}(w^{2}\epsilon_{0}-\bm{k_{g}}(\delta_{z}^{(m)})\bm{\otimes k_{g}}(\delta_{z}^{(m)}))/(\epsilon_{0}T_{g}(\delta_{z}^{(m)}))$, $k_{z}^{(m)}=k_{0z}+\delta_{z}^{(m)}$, $l$ is a thickness of the crystal slab. Again, one should take projections of Eqs.(\ref{eq:eqs6}) on polarization vectors from the analogous problem of incoherent PXR.  One may notice that the boundary problem for incoherent PXR can be formulated in a similar way, namely, by substitution $\bm{E}_{9}\rightarrow\bm{E^{i}}, \delta_{z}^{(9)}\rightarrow 0$, $\bm{E^{i}}$ - inhomogeneous field in case of incoherent PXR. Furthermore, now we will see that $\bm{E}_{9}\approx\bm{ E^{i}}$.
  From Eqs.(\ref{eq:eqs3}) one has the following relation for $\bm{E_{9}}$ (it is valid for any $\bm{E}_{m}$ at corresponding $\delta_{z}^{(m)}$)
 \begin{align}\label{eq:eq7}
 \bm{\widehat{D}}\bm{E_{9}}=\bm{\widehat{J} E_{9}}
 \end{align}
  ,which follows from Eqs.(\ref{eq:eqs1}).
 \begin{align}
 \bm{\widehat{D}}=(T-\bm{k\otimes k})-\frac{W}{T_{g}}\left(\bm{1}-\frac{\bm{k_{g}\otimes k_{g}}}{w^{2}\epsilon_{0}}\right)
 \end{align}
 In Eq.(\ref{eq:eq7}) $\bm{\widehat{D}}$ is calculated at $\delta_{z}=\delta_{z}^{(9)}$. Actually, one can put $\delta_{z}=0$ in $\bm{\widehat{D}}$ and still preserve a good accuracy as $\delta_{z}^{(9)}$ is much smaller than $k_{z0}$ and diffraction roots (by magnitude). But, of course, we save $\delta_{z}=\delta_{z}^{(9)}$ in denominator in $\bm{\widehat{J}}$.

 Similarly, $\bm{E^{i}}$ is the solution of equation
 \begin{align}
 \bm{\widehat{D}}\bm{E^{i}}=\bm{J_{0}}
 \end{align}
\noindent here $\bm{\widehat{D}}$ is calculated at $\delta_{z}=0$.
Together with Eq.(\ref{eq:eq5}) one has the following set of equations
\begin{align}\label{eq:eqs8}
\begin{aligned}
&\bm{\widehat{D}}\bm{E_{9}}\approx\bm{\widehat{J} E_{9}} &&\bm{\widehat{D}}\bm{E^{i}}=\bm{J_{0}} &&&\bm{\widehat{J}E_{9}}=\bm{J_{0}}
\end{aligned}
\end{align}
\noindent here $\bm{\widehat{D}}$ is calculated at $\delta_{z}=0$. In Eqs.(\ref{eq:eqs8}) it's clear that $\bm{E}_{9}\approx\bm{E^{i}}$.
Thereby, we've developed the following algorithm for formulation of boundary condition problem for PXR with bunch instability. One has to write down equations for boundary condition problem in case of incoherent PXR and then replace $\bm{E^{i}}$ in the equations corresponding to the exit surface with $\bm{E^{i}}\exp{(i \delta_{z}^{(9)}l)}$. Applying this approach it is obvious that in the limit $\delta_{z}^{(9)}\rightarrow 0$ or(and) $l\rightarrow 0$ one obtains incoherent PXR.

Eventually, if we consider a thick crystal plate($l>>L_{BraggExt}$), then the equation system (\ref{eq:eqs6}) splits in two independent systems(just like in case of incoherent PXR or X-ray diffraction). One for the entry surface, where one can neglect the specularly  reflected waves, and the other for the exit surface, where the direct waves are negligible. The former system becomes identical to that in case of incoherent PXR, hence,  $\bm{R}$ and $\bm{G_{u}}$ do not encounter any growth. The latter system is similar to that in case of incoherent PXR, but $\bm{E^{i}}$ is replaced with $\bm{E^{i}}\exp{(i \delta_{z}^{(9)}l)}$ now. Hence, the magnitudes of $\bm{T}$ and $\bm{G_{d}}$ in case of PXR with bunch instability are by a factor of $|\exp{(i \delta_{z}^{(9)}l)}|$ bigger than in case of incoherent PXR. Thus, there is an exponential growth.

Now we can name a reason for considering grazing incidence geometry. We have just proved that essentially the enhancement is defined solely by $|\exp{(i \delta_{z}^{(9)}l)}|$. That is, geometry affects the growth through nothing but Cherenkov root $\delta_{z}^{(9)}$. In dispersion equation (\ref{eq:dispeq}) or (\ref{eq:eq11}) Cherenkov root is hidden in the term $(w-\bm{k\cdot\beta})=-\beta_{z}\delta_{z}$. It is clear that as $\beta_{z}$ decreases $\delta_{z}$ increases. And it is grazing incidence geometry that corresponds to small $\beta_{z}$. As one changes geometry the other parts of  Eq.(\ref{eq:dispeq}) or (\ref{eq:eq11}) are changed too, but the above effect proved to be much more significant.

In case of two Cherenkov roots the main logic remains essentially the same. But now one has to replace $\bm{E^{i}}$ with a certain combination $\exp{(i\delta_{z}^{(9)}l)}\bm{E_{9}}+\exp{(i\delta_{z}^{(10)}l)}\bm{E_{10}}$, where $\bm{E_{9}}+\bm{E_{10}}=\bm{E^{i}}$. Exact magnitudes of $\bm{E_{9}}$ and $\bm{E_{10}}$ are unknown yet. They can be found by virtue of the equation for continuity of the electron beam charge density, see \cite{6}. Routine calculations show that there should be the following substitution
\begin{align}
 \bm{E^{i}}\rightarrow \left(\exp{(i\delta_{z}^{(9)}l)}\frac{\delta_{z}^{(9)}}{\delta_{z}^{(9)}-\delta_{z}^{(10)}}+\exp{(i\delta_{z}^{(10)}l)}\frac{-\delta_{z}^{(10)}}{\delta_{z}^{(9)}-\delta_{z}^{(10)}}\right)\bm{E^{i}}
\end{align}
\section{Numerical results}  \label{sec:sne}
\begin{figure}
  \centering
  \includegraphics[width=0.95\textwidth]{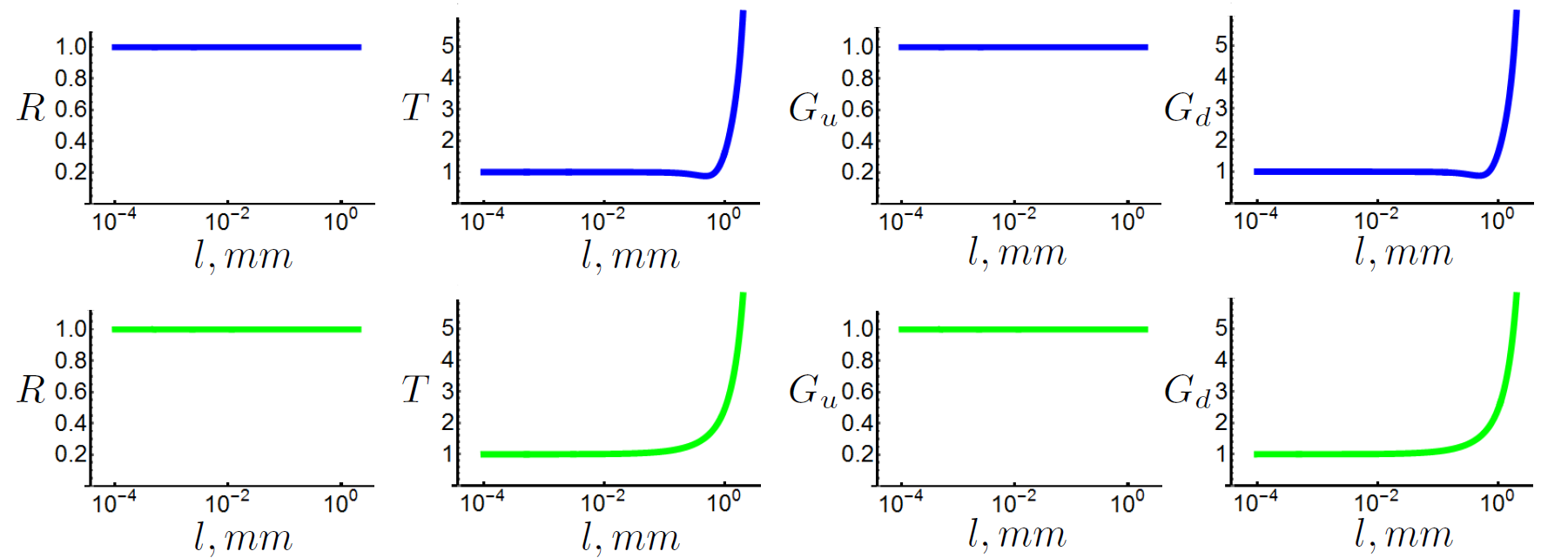}\\
  \caption{Dependence of amplification with respect to incoherent PXR on thickness of the crystal slab. Blue plots correspond to zero emittance case, green - to nonzero emittance case (blue and green plots are obtained at different beam current densities).}\label{fig:figure5}
\end{figure}

 Fig.\ref{fig:figure5} presents numerical calculation of amplitude amplification obtained by accurately formulated boundary condition problem.

\begin{figure}
  \centering
  \includegraphics[width=0.7\textwidth]{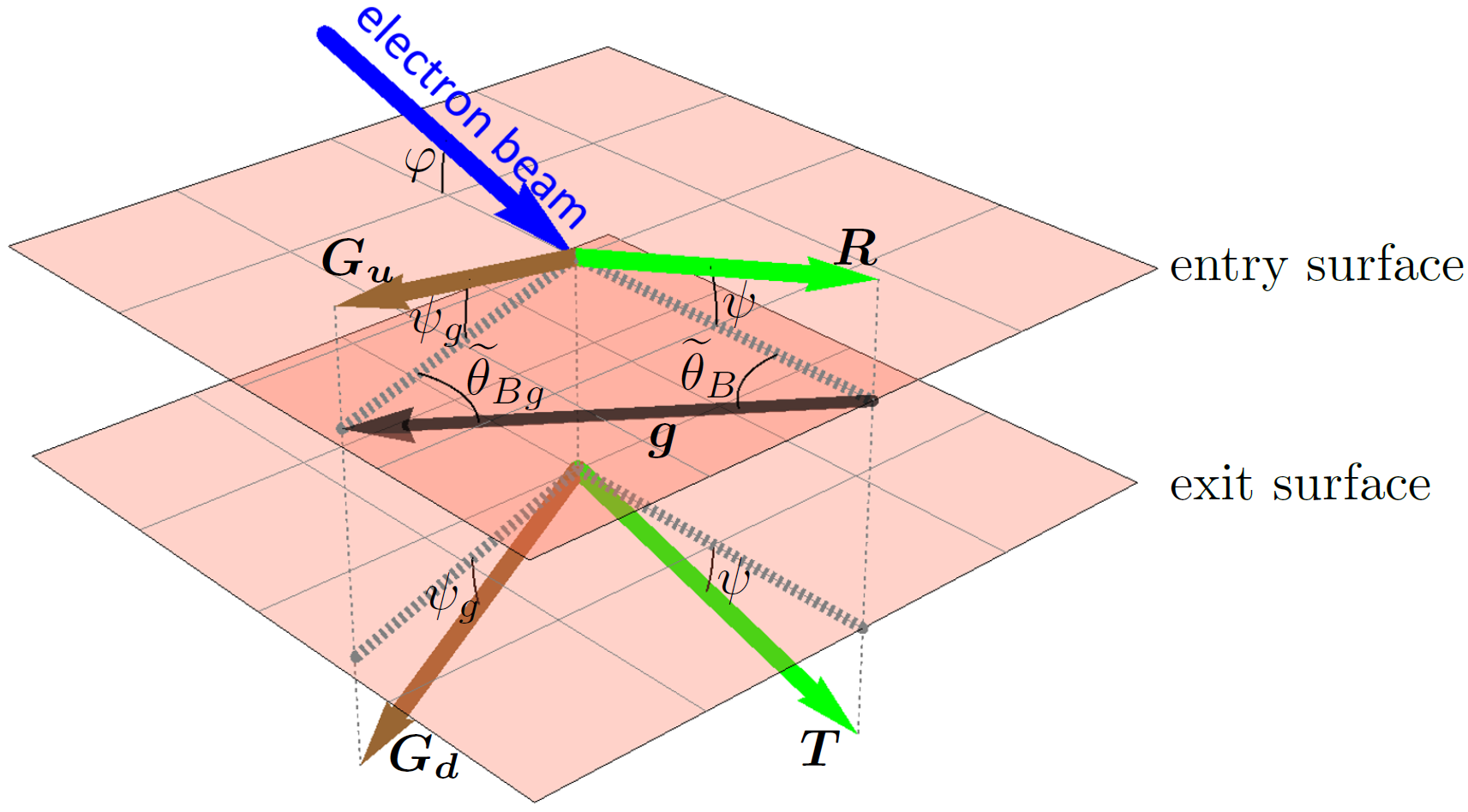}\\
  \caption{Relative orientation of the wavevectors and the parameters defining it? supplementary to as Fig.\ref{fig:figure3}. Here $\widetilde{\theta}_{B}=\pi/2-\theta_{B}$, $\widetilde{\theta}_{Bg}=\pi/2-\theta_{Bg}$.}\label{fig:figure6}
\end{figure}

The Si crystal was considered, surface normal is $(0,0,1)$ and is parallel to $z$-axis ($\bm{n}$), $(h,k,l)=(0,4,0)$. The parameters describing geometrical configuration in case of zero emittance are shown in Fig.\ref{fig:figure6}. Their numerical values are  $\varphi=\text{1${}^{\circ}$27'44.061$\texttt{"}$}$, $\theta_{B}=\text{30${}^{\circ}$0'30.193$\texttt{"}$}$, $\theta_{Bg}=\text{30${}^{\circ}$0'0.629$\texttt{"}$}$, $2\pi c/w=0.13578522 nm$, $\psi=\text{0${}^{\circ}$17'22.336$\texttt{"}$}$, $\psi_{g}=\text{0${}^{\circ}$47'31.108$\texttt{"}$}$, the beam current density $j_{\epsilon=0}=1.25\times 10^{9}A/cm^{2}$.

%
%

In finite emittance case, geometrical configuration is like in zero emittance case except for the direction of the beam electrons' velocities, in both cases the beam electrons' energy is $E_{e}=100MeV$. In finite emittance case we have to set borders of the distribution over directions. Here we used $\beta_{zmin}=6.34\times 10^{-3}$, $\beta_{zmax}=8.07\times 10^{-3}$, $X_{min}/w=0$, $X_{max}/w=3.75\times 10^{-5}$, see Sec.\ref{sec:hpoebc}.The normalized emittance corresponding to these values is $\gamma \epsilon=1.42\times 10^{-7}m rad$ assuming focusing to $0.1 \mu m^2$, the peak current was taken as $I=10 kA$; the emittance and peak current values are on the frontier of achievable values (see \cite{14,15,16}). If one uses approximations similar to (\ref{eq:j_beta_otimes_beta}) and analogue of (\ref{eq:dispeq}), one can find that the increment depends on ratio $I/\epsilon^2$, i.e. beam brightness. Thereby, to merely check existence of the effect on practice one should use an electron beam with calculated emittance (or less) and peak current and direct it so that its mean $\beta_{z}$ lies within the interval $(\beta_{zmin},\beta_{zmax})$. Then, one can expect amplified(with respect to incoherent PXR) amplitudes of the waves leaving the crystal through the exit surface in directions described in the above paragraph. 

 \begin{figure}
  \centering
  \includegraphics[width=0.8\textwidth]{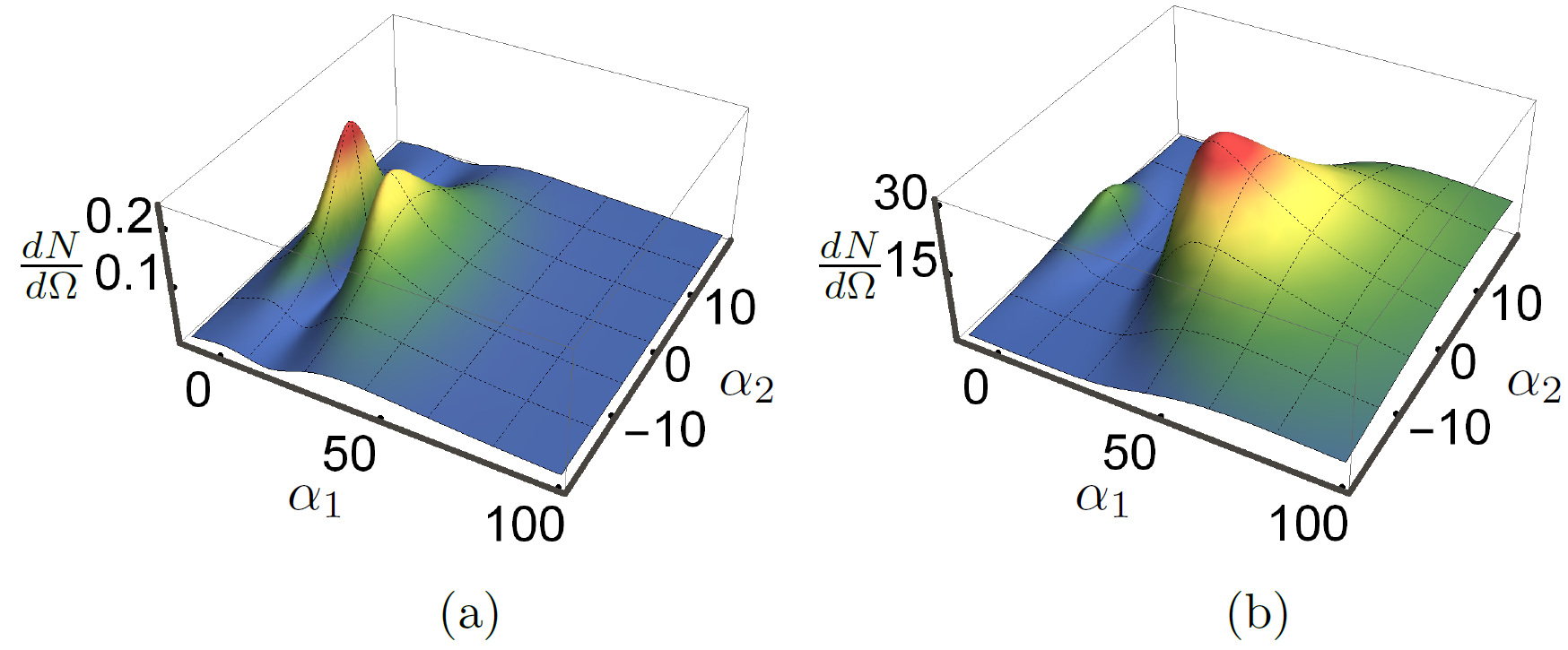}\\
  \caption{Angular distribution of emitted by one electron quanta ($N$ is a number of photons, $\Omega$ is a solid angle) for incoherent PXR (a) and PXR with beam instability (b). $\alpha_{1}$ and $\alpha_{2}$ are measured in $mrad$. $\bm{G_{d}}$-wave is considered.}\label{fig:figure8}
\end{figure}

%

%


In Fig.\ref{fig:figure8} shows angular distribution of emitted quanta and comparison with that under incoherent PXR, the case of zero emittance is considered. The same parameters as in Fig.\ref{fig:figure6} are used except for thickness of the crystal slab $l$ which is increased to $3mm$.  $\alpha_{1}$ is an increase in $\psi_{g}$ from Fig.\ref{fig:figure6}, $\alpha_{2}$ is a decrease in $\widetilde{\theta}_{Bg}$. For considered region in $\alpha_{1},\alpha_{2}$ we performed an integration and obtained the following estimations for photon yield. $1.5\times 10^{-4} photons/electron$ in case of incoherent PXR and $3.7\times 10^{-2}photons/electron$ in case of PXR with beam instability. Obtained value for photon yield of incoherent PXR (in grazing geometry) is of the same order as in \cite{22},  for details of PXR under condition of grazing incidence diffraction see also \cite{23,24}.

\section{Discussion and conclusions}\label{sec:discussion}
Let us now return to Fig.\ref{fig:figure5} and discuss it. Firstly, we see that the waves leaving the crystal through the entry surface do not undergo amplification at all. It is in agreement with predictions of the approximate solution of boundary condition problem.

Secondly, we chose such parameters that amplification is of the same order in both cases on purpose. This way we see that the character of the shift to the regime of exponential growth is slightly different. The plot is sharper in zero emittance case. In nonzero emittance case there is only one Cherenkov root and one term $\exp(i \delta_{z}^{(9)}l)$ in equations, which provides steady and smooth increase. On the contrary, in zero emittance case there is a struggle between two Cherenkov roots, see Eq.(\ref{eq:eq15}), one from which provides increase, while the other provides decrease.  That is why there is longer plateau, short depression and sharper increase in the end, where one root finally overpowers the other. However, at larger energies the depression disappears, but the effect as a whole slowly diminishes.

Further, after $l=1mm$ amplification exponentially increases and tends to infinity, but one has to remember that it is a linear theory of the effect. It describes solely the startup process. In order to adequately describe further evolution one should apply more exact non-linear theory. It will result in saturation, when the intensity becomes constant and proportional to $N^{2}$, where $N$ is a number of electrons in the beam, instead of $N$ in case of incoherent PXR. That is, the electron beam starts to emit as a single charge. The calculated photon yield for PXR with beam instability ($3.7\times 10^{-2}photons/electron$) increased by two orders with respect to incoherent PXR ($1.5\times 10^{-4} photons/electron$). That is much less than the number of electrons in the beam $N$ and corresponds to the startup of the process far from saturation. It makes the estimation for photon yield reliable. 

As to obtained angular distributions of emitted photons, then in Fig.\ref{fig:figure8}a one can recognize a well-known pair of incoherent PXR peaks (see \cite[p.31]{1}), although in this case they are asymmetric due to peculiar geometry (grazing incidence). In Fig.\ref{fig:figure8}b we see that the peak, corresponding to higher angles to the crystal's surface, undergoes stronger amplification.

In present contribution we allowed for absorption in the crystal and showed that it considerably reduces the increment of instability ($-\operatorname{Im}(\delta_{z})$).  In zero emittance case $\delta_{z}\sim j^{1/2}$ ($j$ is the beam current density), in nonzero emittance case $\delta_{z}\sim j$. For comparison, under vanishing absorption $\delta_{z}\sim j^{1/3}$, this case is studied in detail in \cite[pp.146-153]{1}.

In nonzero emittance case we presented results for electron beam with constant emittance. However, as the beam passes through the crystal its properties change, the most considerable influence being the Coulomb scatter on the crystal atoms' nuclei. If one takes into account corresponding increase in emittance during passing through the crystal, then the effect is essentially suppressed. We figure that the effect requires the beam to be restrained from expanding direction-wise. This may be fulfilled to a certain extent by channeling of the beam electrons. This way, considerable part of the beam electrons travel along the crystal within a definite range of directions, characterized by Lindhard angle $\Theta_{L}$ (see \cite{8},\cite[p.32]{9}). In course of time the amount of electrons in channeling regime descends. Still, if, for instance, the amount decreases as $1/\sqrt{z}$ (see \cite{12,13}), one can expect an integral gain proportional to $\sqrt{l}$ (that is, amplification $\sim \exp(const\times\sqrt{l})$), where $z$ is the in-crystal coordinate, $l$ is the crystal plate's thickness. The beam instability in combination with channeling of the electrons will be considered in further research.

One more possible improvement is to use beforehand nano-modulated electron beam. In \cite{14} a new method of obtaining such electron beams is described. Nano-modulation does not change the increment of instability, but it permits to reduce the length of the plateau in Fig.\ref{fig:figure5}. In this case fast exponential growth of intensity may start from smaller crystal thicknesses, that also reduces the Coulomb scatter negative effects.

\bibliographystyle{unsrt}

\end{document}